\documentclass[a4paper, 11pt]{article}
\usepackage[utf8]{inputenc}
\usepackage{graphicx}
\usepackage{hyperref}
\usepackage{url}
\usepackage{geometry}
\usepackage{fancyhdr}
\usepackage{caption}

\title{FUTURAL: A Metasearch Platform for Empowering Rural Areas with Smart Solutions}
\author{
    Matei Popovici\\
    \texttt{matei.popovici@upb.ro}
    \and
    Ciprian Dobre\\
    \texttt{ciprian.dobre@upb.ro}
}
\date{May 2024}

\begin{document}

\maketitle

\begin{abstract}
\noindent The FUTURAL project aims to provide a comprehensive suite of digital Smart Solutions (SS) across five critical domains to address pressing social and environmental issues. Central to this initiative is a robust Metasearch platform, which will not only serve as the primary access point to FUTURAL's solutions but also facilitate the search and retrieval of SS developed by other initiatives. This paper elaborates on the MVP implementation for the MetaSearch platform. It focuses on a single, open-source data service and harnesses the generative capabilities of Large Language Models (LLMs) to create a user-friendly natural language interface. The design of the Minimum Viable Product (MVP), the tools used for adapting LLMs to our specific application, and our comprehensive set of evaluation techniques are thoroughly detailed. The results from our evaluations demonstrate that our approach is highly effective and can be efficiently implemented in future iterations of the MVP. This groundwork paves the way for extending the platform to include additional services and diverse data sets from the FUTURAL project, enhancing its capacity to address a broader array of queries and datasets.
\end{abstract}

\clearpage
\tableofcontents
\clearpage

\section{Introduction}

The digital transformation of rural areas is a critical step towards addressing pressing social and environmental issues. The FUTURAL project is at the forefront of this transformation, aiming to provide a comprehensive suite of digital Smart Solutions (SS) across five critical domains. A central component of this initiative is a robust Metasearch (MS) platform, designed to be the primary access point to FUTURAL's solutions and to facilitate the search and retrieval of SS developed by other initiatives.

\subsection{Motivation and Problem Statement}
Rural communities often face challenges in accessing and utilizing digital technologies. While specialized information, such as weather forecasts for agriculture, is increasingly available online, the interfaces to access this information are often complex and not user-friendly for individuals with limited IT experience. There is a clear need for a platform that can bridge this gap by providing a simple, intuitive, and natural language-based interface to a wide range of smart services.

The problem we address is twofold: first, the integration of heterogeneous data from various service providers, and second, the presentation of this data to the user in a coherent and easily understandable format. The MS platform is designed to tackle these challenges by leveraging the power of Large Language Models (LLMs).

\subsection{The FUTURAL Project}
The FUTURAL project is a HORIZON Innovation Action funded by the European Union. Its main objective is to empower rural areas through innovative smart solutions. The project brings together a consortium of 21 partners from across Europe, with expertise in various fields, including agriculture, technology, and social sciences. The project started on June 1, 2023, and will run for 48 months.

\subsection{MVP alignment with the MS platform architecture}
In this section we briefly review the MS (MetaSearch) platform architecture. The architecture is illustrated in Figure \ref{fig:fig1}, and consists of (i) the input processing module, which analyses user queries, extracts semantic information from such queries and makes decisions regarding the appropriate Services that should be queried; (ii) the Metadata Management Module, which is responsible for holding information specific to each Service Provider that will be available on the MS platform as well as information specific to the type of data available, and the format in which it is available; (iii) the Data Service Integration module maps the user search to a set of actual API-specific queries that will be performed at each appropriate Service Provider and (iv) the Semantic Data Processing module, which will transform the query results, from each format specific to each Service Provider, to a coherent, unified, human-readable format which combines text with images as well as other human-readable artefacts.

The focus in the first MVP implementation for the MS platform will be on modules (i), (iii) and (iv), with a particular emphasis on the Semantic Data Processing Module. This module relies on the usage of Large Language Models (LLMs), and requires adaptation of specific LLMs that have already been developed, as well as LLM-based techniques to the needs of the MS platform. Our MVP implementation and its relation to the aforementioned modules is detailed in the following sections.

\section{Related Work}

\subsection{Metasearch Platforms and Heterogeneous Data}
Metasearch engines traditionally aggregate results from multiple web search engines. However, the FUTURAL platform addresses a more complex challenge: integrating heterogeneous data from a variety of Smart Services, each with its own API and data format. This requires a more sophisticated approach than traditional metasearch. Recent research has explored the use of LLMs to tackle the challenges of heterogeneous data integration. For instance, LLMs can be used for tasks like entity matching, schema alignment, and wrapper induction to create a unified view of disparate data sources \cite{vldb2023, arxiv2023}. Our work builds on these ideas, using an LLM as the core component for both data integration and user interaction.

\subsection{Natural Language Interfaces for Data Retrieval}
The idea of using natural language to query databases dates back several decades. However, early systems were often brittle and limited to a narrow range of queries. The recent success of LLMs has renewed interest in Natural Language Interfaces (NLIs). Modern NLIs leverage the power of LLMs to understand complex queries, handle ambiguity, and generate responses in natural language. For example, systems like an NLI for SQL (NLISQL) can translate natural language questions into SQL queries, making databases more accessible to non-technical users. Our approach is similar in spirit, but we go beyond simple query translation by using the LLM to also process the retrieved data and generate a comprehensive, human-readable summary.

\subsection{LLM Fine-tuning for Domain-Specific Tasks}
While large-scale pre-trained LLMs possess a vast amount of general knowledge, they often lack the domain-specific expertise required for specialized tasks. Fine-tuning is a common technique to adapt pre-trained LLMs to a specific domain. For example, models like SciBERT and BioBERT have been fine-tuned on scientific and biomedical text, respectively, achieving state-of-the-art results on various NLP tasks in these domains. In our work, we fine-tune an open-source LLM on a dataset of weather forecasts to improve its ability to generate accurate and fluent weather reports.

\subsection{Retrieval-Augmented Generation (RAG)}
A key challenge for LLMs is the problem of "hallucination," where the model generates plausible but factually incorrect information. Retrieval-Augmented Generation (RAG) is a promising technique to address this issue. RAG combines a generative LLM with a retrieval system that can access an external knowledge source, such as a document collection or a database. When a user asks a query, the retrieval system first finds relevant information from the knowledge source, and then the LLM uses this information to generate a response. This ensures that the generated response is grounded in facts. Our future work plans include incorporating a RAG-based approach to allow the MS platform to access a wider range of data sources and provide more accurate and reliable answers.

\section{System Architecture}

\begin{figure}[htbp]
    \centering
    \includegraphics[trim={1cm 1cm 16cm 1cm}, clip, width=0.5\textwidth]{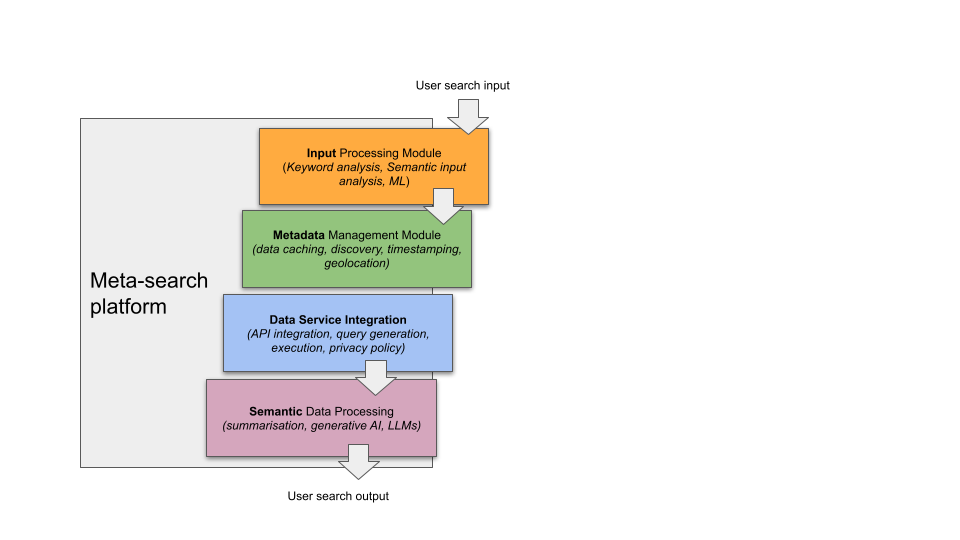}
    \caption{Overall MetaSearch platform components.}
    \label{fig:fig1}
\end{figure}

The MS platform architecture is designed to be modular and scalable, encompassing a comprehensive design for data governance and interoperability. As illustrated in Figure \ref{fig:fig1}, the architecture consists of four main modules.

\subsection{Input Processing Module}
This module is the entry point for user queries. It is responsible for analyzing the query, extracting semantic information, and making decisions about the appropriate services to query. In the final version of the platform, this module will use a combination of keyword analysis, semantic analysis, and machine learning techniques to understand the user's intent.

\subsection{Metadata Management Module}
The Metadata Management Module holds information about each service provider available on the platform. This includes details about the type of data available, its format, and the API endpoints. This module is crucial for the dynamic integration of new services and for ensuring the interoperability of the platform.

\subsection{Data Service Integration Module}
This module maps the user's search to a set of API-specific queries that will be performed at each appropriate service provider. It is responsible for handling the communication with the external services, including authentication and error handling.

\subsection{Semantic Data Processing Module}
The Semantic Data Processing module is the core of our LLM-based approach. It transforms the query results from each service provider's specific format into a coherent, unified, and human-readable format. This module combines text, images, and other artifacts to generate a comprehensive response to the user's query.

\section{MVP Implementation}

\subsection{Overview}
The MVP implementation aims to quickly prototype our MS platform design ideas using LLMs. To achieve this, we focused on leveraging readily accessible third-party data from the project's outset, avoiding delays associated with waiting for specific data to be available. Our MVP is a stand-alone application that uses public weather API data in JSON format to create a human-readable weather forecast. The main use-case is as follows: A user searches by asking a natural language question about upcoming weather conditions in a specific location and for a given timeframe (e.g., the next day, the next 7 days, or the next 14 days). The MVP processes the query to extract the location and timeframe, accesses relevant data from the API service, which is available in tabular form, and contains precipitation probabilities, precipitation amounts per hour, wind directions and speeds, etc. This data is retrieved in JSON format, which we feed to our Large Language Model. Our LLM then generates the desired human-readable forecast.

We believe the MVP can function as a readily deployable smart service on its own. In many countries of the EU, especially Romania, specialised weather forecasts are available and Internet connectivity is becoming the norm rather than the exception for many rural locations. However, many of the people that are particularly interested in these accurate forecasts for agricultural work are not fluent users of the underlying technologies. Installing apps and navigating them, as well as getting acquainted with technical interfaces is not straightforward for many people in rural communities, especially since these apps have features of ever-increasing complexity (satellite \& radar projections, multi-model forecasts, etc). Thus, being able to distil such complex information into a human-readable yet accurate and specific text weather forecast is, and will continue to be for a long time, a very useful feature for such apps.

\begin{figure}[htbp]
    \centering
    \includegraphics[trim={1cm 4cm 8cm 0cm}, clip, width=0.8\textwidth]{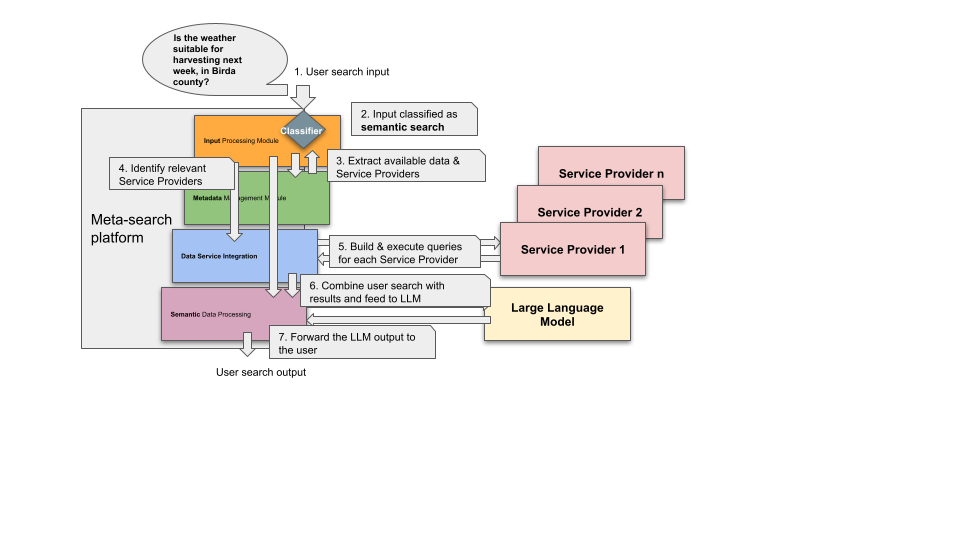}
    \caption{The first MVP implementation mapped on the MS platform architecture.}
    \label{fig:fig2}
\end{figure}

Figure \ref{fig:fig2} illustrates how our MVP implementation and use case maps to the MS platform architecture from Figure \ref{fig:fig1}. The user query will be classified by the Input Processing Module. In this first MVP, input processing (point 2.) will be minimal and focus solely on location and timeline extraction. Our MVP utilises a single service (point 4.), so metadata considerations are not relevant at this stage. The forecast API, our sole service, will be accessed with a specified location and timeline (point 5.). The JSON-formatted response, along with the original user query, is then fed into our Large Language Model (point 6.), which generates a weather forecast. This forecast text is then presented to the user. (point 7.)

\subsection{MVP Deployment}
\begin{figure}[htbp]
    \centering
    \includegraphics[trim={1cm 5cm 1cm 3cm}, clip, width=0.8\textwidth]{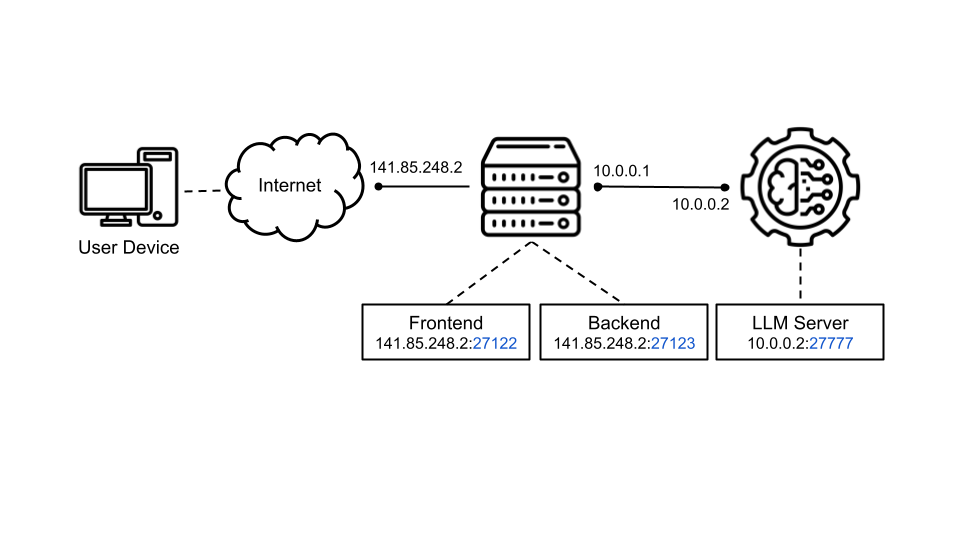}
    \caption{MS platform Deployment Platform.}
    \label{fig:fig3}
\end{figure}

The MS platform MVP is currently deployed as described in Figure \ref{fig:fig3}. We use two machines, one for frontend and backend servers, and another one for LLM processing. This is necessary since running LLMs requires dedicated GPU hardware which is usually either shared between multiple applications, or available as a purchasable resource from third party GPU providers. For security reasons, the MS platform prototype is not publicly accessible 24/7 while it is still in development. This deployment is subject to change as we improve the MVP in the next iteration with better scalability solutions.

The two servers are running as Docker containers. The frontend server is responsible for the MVP user interface, as well as for forwarding of user input to the backend server. The backend server is responsible for processing user input and sending it to the LLM server, as well as receiving the LLM response and sending it back to the frontend server. It ensures that the input is properly formatted and performs input sanitization to prevent security issues.

Because we do not store client messages on the server, even without input sanitization, the current state of the client-side web interface cannot introduce vulnerabilities such as Cross Site Scripting, where malicious code is stored on our servers and subsequently reflected back to other users, nor can it lead to unauthorised modifications of data on the server. However, this does not inherently protect against all types of client-side issues such as CSS Injection. CSS Injection can still occur if user inputs are dynamically used in styles and are not properly sanitised, which could potentially degrade the user experience by altering the visual presentation of the application. While these issues do not compromise server security directly or facilitate the theft of sensitive data via server mechanisms, they can still pose risks to the integrity of user interactions and data privacy on the client side.

The backend server supports the following two URIs:
\begin{itemize}
    \item \textbf{/html} $\rightarrow$ \textbf{POST requests} to this URI are used to generate HTML code for the chat messages. The request body is a JSON object with the following fields: sender (string), text (string), and colour (string). The response is a JSON object with the field message containing the generated HTML code for the user interface. There are two possible responses: 200 (OK) and 500 (Internal Server Error).
    \item \textbf{/meteo-query} $\rightarrow$ \textbf{POST requests} to this URI are used to send user input to the LLM server. The request body is a JSON object with the field message (string). The response is a JSON object with the field message containing the LLM response. There are two possible responses: 200 (OK) and 500 (Internal Server Error).
\end{itemize}

The second machine, running the LLM server, is equipped with the following GPU setup: 3x NVIDIA AD102 GeForce RTX 4090 with an availability of 24576 MB VRAM each. The GPUs are used to accelerate LLM processing, which is computationally intensive. The GPUs are accessed through the CUDA API, which allows for parallel processing of the LLM.

The LLM server is responsible for receiving queries from the backend server, processing them using the LLM, and sending the response back to the backend server, while also logging the queries and responses for debugging purposes. The server always validates and sanitises the input before processing. The LLM server supports the following URI:
\begin{itemize}
    \item \textbf{/meteo} $\rightarrow$ \textbf{POST requests} to this URI are used to send user input to the LLM model. The request body is a JSON object with the field message (string). The response is a JSON object containing the LLM response. There are two possible responses: 200 (OK) and 500 (Internal Server Error).
\end{itemize}

\subsection{Input Processing and Querying}
In the MVP implementation, the Input processing module will not perform Smart Service classification, since we only use a single data-source. Instead, this module will detect information related to user location for generating the forecast. Detection proceeds as follows. The frontend server receives the user’s query and forwards it to the backend server, which processes the message and sends it to the LLM server. The LLM server processes the message and returns the response to the backend server, which then forwards it to the frontend server. The frontend server displays the response in the chat window. The user can continue the conversation by sending additional messages.

The LLM server first processes the query to identify if it is related to a weather forecast. If so, it employs Named Entity Recognition (NER) to extract the city names mentioned in the query, which are then matched against a local database containing latitude and longitude coordinates to accurately locate the city. For time references, the model uses NER to identify time mentions. However, in this first MVP, we assume the time is "today," since the current data provider, Meteoblue, only offers weather data for the current day.

Once the location and assumed time are determined, the system uses these details to query the Meteoblue weather API with the geographic coordinates. The API responds with a weather forecast in the form of a JSON object, which is then fed into the LLM model. The LLM model processes this data to generate a contextually relevant weather forecast response. This response is then sent back to the user via the backend server.

If the query is unrelated to weather forecasts, the LLM server generates a response based on the content of the query itself, without invoking third-party service providers. It is important to note that in the current stage of development, we do not provide the LLM model with the context of previous messages. Therefore, responses are generated based solely on the current message, without taking into account the chat history.

For our weather API, we collect the data from Meteoblue.com, by querying pages with the URI format \url{https://www.meteoblue.com/en/weather/week/<latitude>N_<longitude>E} where we replace \texttt{<latitude>} and \texttt{<longitude>} with the corresponding values, using a precision of 3 decimal places. We then parse the HTML content of the page to extract the JSON data and the weather bulletin text.

\begin{figure}[htbp]
    \centering
    \includegraphics[trim={1cm 2cm 1cm 1cm}, clip, width=0.8\textwidth]{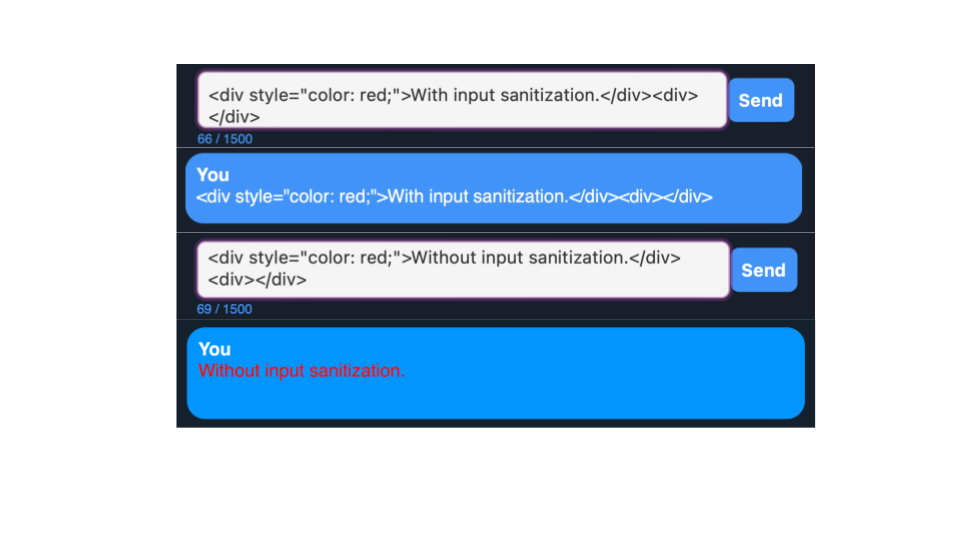}
    \caption{Effects of input sanitization in the chat interface.}
    \label{fig:fig4}
\end{figure}

\subsection{Generating User Answers using LLMs}
This section outlines our step-by-step design process for creating the Large Language Model (LLM) that we use to generate forecasts.
Existing Large Language Models like GPT-3.5 perform reasonably well when prompted with weather forecasts in tabular formats and human queries about the weather. However, this approach has limitations: (i) it doesn't scale with the vast amount of data expected from the FUTURAL project, and (ii) it relies on proprietary technology (GPT models) that is not open-source and cannot be adapted to our needs. For this reason, our approach for the implementation of our LLM focuses on selecting an appropriate open-source LLM, and deploying adaptation strategies to increase the performance of the chosen LLM to our forecast generation task. In what follows we briefly review candidates for a suitable LLM as well as adaptation strategies that we have considered in implementing our LLM for the MVP.

\subsubsection{Available options for selecting an open-source LLM}
Mistral \cite{mistral} models vary in size and computational requirements:
\begin{itemize}
    \item \textbf{Mistral 7B} - is a dense model for quick experimentation and iteration containing 7 billion parameters. It is known to match 30B parameter models in performance, with a context size of 32k tokens;
    \item \textbf{Mistral 8x7B} - is a sparse model utilising roughly 12B out of 45B parameters for enhanced throughput, featuring a context size of 32k tokens;
    \item \textbf{Mixtral 8x22B} - is a larger sparse model using about 39B of its 141B parameters, designed for better throughput and increased vRAM demand, with a context size of 64k tokens.
\end{itemize}

The Mistral family of models have function-calling \cite{mistral_function_calling} abilities that allow for the execution of calling functions from within the model, enabling the use of external APIs and services, however our experiments have found these to be limited in their performance. We opt for implementing our own function calling system that allows for the execution of external APIs and services using more accessible and more well documented models such as Llama-2 and Llama-3.

Llama-2 \cite{llama2} is a family of LLMs that allow a context length of 4096 tokens, which is about 3150 words, with an average of 1.3 tokens per word in English. For the chat interface of the MVP implementation, we opted for the fine-tuned Llama-2-Chat model, optimised through Reinforcement Learning from Human Feedback for dialoguing with users. It is a general-purpose LLM, meaning that it can be used for a wide range of tasks, including chat, question-answering, and text generation. We currently utilise the 7 billion (7B) parameters model, the smallest in the Llama-2 series, for prototyping. This model offers advantages in memory usage and computational efficiency, facilitating quicker and more cost-effective development of prototypes. Nonetheless, it does not achieve the accuracy levels of its larger counterparts, the Llama-2-13B and Llama-2-70B models, which contain 13 billion and 70 billion parameters respectively and require significantly more memory and computational power. We chose the 7B model for our initial prototype to strike a balance between accuracy and development efficiency, with the option to upgrade to a more powerful model should the need for greater accuracy arise in future iterations. For the current prototype, we illustrate adaptation techniques that increase the performance of this model choice for our forecast generation task.

The Llama-3 family is a newly launched family of LLMs (as of April 2024), with two variants available so far: 8B and 70B, both options having instruction tuned versions that can be used for chat applications. With a context size of 8k tokens, the Llama-3 models are designed to be more efficient and faster than the previous Llama-2 models. We aim to investigate the capabilities of the Llama-3 models in future stages of the project.

\subsubsection{Overview of LLM adaptation techniques for the MS platform}
Large Language Models (LLMs) rely on the transformer \cite{attention} architecture and usually consist of parameters in the order-scale of billions. This parameter scale gives them the ability to recognize language patterns and grasp the meaning of input text. In order to achieve high performance for any language-related tasks, LLMs gain their language comprehension abilities by learning semantic relations between words. These relations are learned by seeing vast amounts of human-like texts, either taken from available sources online or specially curated for a specific task.

Pretraining a large language model is a task achieved using a massive corpus of text data, in the order scale comparable to the text available on Wikipedia. The gathered text is split into words or subwords named tokens which are then converted into a vectorial representation. The overall objective of pretraining is to obtain a model that, given a sequence of tokens as input, will predict the most likely next-word in a given sequence, with respect to the large amount of text that the model has seen. Pretraining a model from scratch for the MS platform is unfeasible, for two main reasons: (i) the sheer amount of data required for this task; (ii) the computational effort and hardware requirements for pretraining, which is prohibitively expensive. Fortunately, we can build LLMs appropriate for the MS platform by other means.

A useful particularity of transformer-based LLMs is their ability to be customised via transfer-learning \cite{transfer_learning}. Transfer learning is a technique that allows a model developed for a particular task to be reused as the starting point for creating a model for a second task. It is particularly useful when the second task has limited training data. To achieve this, a model is split in two parts: a task-specific head and a general-purpose body. The body learns to recognize language patterns and the relations between words by training on a large-scale dataset, as illustrated in the previous section. Then, the model is fine-tuned by making modifications to the layers that form the head, through training on a task-specific dataset. The models obtained this way have a high generalisation capacity and can be easily be repurposed and used in various applications.

This process leverages the model's intrinsic knowledge of language, gained during pre-training, to perform better on the new task. There are various methods of fine-tuning, ranging from updating all the model's weights (full fine-tuning) to training only a small number of additional weights (adapter tuning).

Full fine-tuning involves updating the weights across all layers of the model. Although this approach is computationally intensive and requires significant storage—since it results in multiple copies of the model, one for each task—it tends to yield the most effective outcomes.

\begin{figure}[htbp]
    \centering
    \includegraphics[trim={1cm 6cm 1cm 1cm}, clip, width=0.8\textwidth]{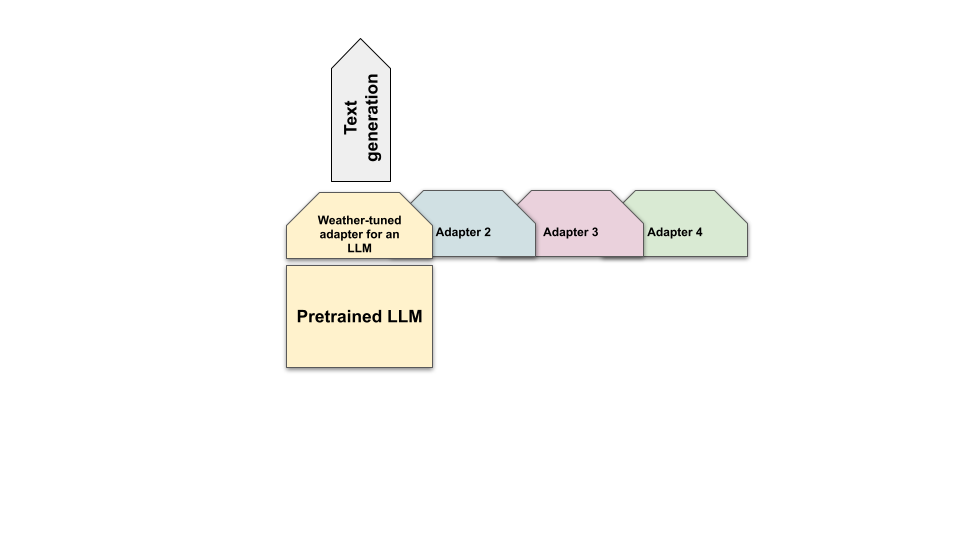}
    \caption{Using adapters to customise a Large Language Model.}
    \label{fig:fig5}
\end{figure}

Alternatively, adapter tuning involves adding small, task-specific neural networks, or adapters, to the model. These adapters are trained while the original model's weights remain frozen. This method is less resource-intensive as it requires less computational power and storage, maintaining only one copy of the baseline model and small adapters for each task. The modular nature of adapters also facilitates easy integration of new tasks without retraining the entire model. This approach offers a practical balance between performance and resource utilisation, making it a viable option for managing multiple tasks efficiently, which will be especially useful for the final version of the MS platform, because it allows us to create multiple adapters (as illustrated in Figure \ref{fig:fig5}) which can be combined in different ways for different tasks.

Lastly, prompt engineering offers another approach to leveraging a pre-trained model for specific tasks without the need for fine-tuning. This method involves providing the model with a few task-specific examples as prompts, allowing it to perform the task by extrapolating from these examples. While very efficient in terms of computational resources, prompt engineering does not always achieve the same level of effectiveness as task-specific fine-tuning, but we are considering it for specific tasks such as query classification, subject to the implementation of Input Processing Module, for future iterations of the MS platform implementation.

\subsubsection{Adapter tuning our model}
\paragraph{Building the training dataset.}
In order to fine-tune Llama-2-7B on the weather bulletin generation task, we have built a dataset of weather data in the format specific to our chosen weather API. The dataset is a collection of JSON inputs (that contain real and felt temperature, wind speed and direction, precipitation size and probability, cloud coverage for every 3 hours in a day), as well as corresponding output, namely the weather bulletin text, which is a description in natural language of the data provided.

To create this dataset, we employed web scraping techniques on the Meteoblue.com website to collect data bulletins. We randomly selected latitude and longitude values to ensure unbiased global coverage, resulting in a diverse distribution of meteorological conditions. Over several days, we scraped the website and amassed a total of 44,550 input-output data pairs.

We have built an evaluation dataset in a similar way, which contains 500 pairs of input JSON and output weather forecasts, in order to evaluate the performance of our fine-tuned model on the weather bulletin generation task.

\paragraph{The Hugging Face API for Fine-Tuning LLMs.}
Hugging Face Transformers \cite{transformers} is a library that provides state-of-the-art NLP models (such as BERT, GPT-2, RoBERTa, etc.) and pipelines. It is built on top of PyTorch and TensorFlow. The library is widely used for a variety of NLP tasks, such as text classification, named entity recognition, and question answering. We use its APIs for model fine-tuning, inference and evaluation. We employed Parameter-Efficient Fine-Tuning (PEFT) \cite{peft} using the Hugging Face Transformers library. PEFT enables the fine-tuning of only a small subset of the model's parameters, keeping the rest fixed, which significantly enhances computational efficiency and allows for effective training on smaller datasets. Specifically, we utilise specialised adapters within PEFT, such as Low-Rank Adapter (LoRA) \cite{lora} and Quantized Low-Rank Adapter (QLoRA) \cite{qlora}. LoRA integrates trainable low-rank matrices into a model's architecture by inserting them in specific layers (like attention or feed-forward layers) to apply focused adjustments that modify the output of the model's existing layers. This means the model gets a bit bigger but is better at handling specific tasks. The fine-tuning process with LoRA is faster and uses less computation power compared to full model fine-tuning. Additionally, QLoRA enhances this approach by adding quantization, which reduces the model's memory use and lowers the computing power needed for fine-tuning. In quantization, the model's parameter values, which are usually continuous, get simplified into smaller, discrete numbers. This means going from more detailed 32-bit numbers to simpler 8-bit numbers. We have experimented with several hyperparameter configurations, including various learning rate values and batch sizes, and have obtained the best results with an Alpha of 16, R of 64, weights quantization type \texttt{nf4}, computation type \texttt{float16}, 1 epoch, batch size 1 and learning rate of 0.0001.

\section{Evaluation}

\subsection{Evaluation Methodology}
Evaluating the performance of Large Language Models for various tasks is often challenging. One source of difficulty stems from the fact that conversational models are non-deterministic - their output can vary from query to query. Furthermore, there is no-ground truth response that is uniquely valid. More specifically, in our context, there is no single weather forecast text that is correct, given a certain data prediction in tabular form. However, we can deploy approximation metrics such as the BLEURT and ROUGE scores coupled with human evaluation and AI-powered evaluation. Using this combination of metrics we can get an accurate assessment of the model's performance.

\subsection{Automated Metrics: ROUGE and BLEURT}
ROUGE (Recall-Oriented Understudy for Gisting Evaluation) \cite{rouge} score, is a set of metrics used to evaluate the quality of text summaries. Using ROUGE, we have compared the output of our LLM with the set of reference summaries that we have built. The comparison is achieved using n-grams, i.e. word sequences of length n. There are three variants of ROUGE:
\begin{itemize}
    \item ROUGE-1 measures the overlap of unigrams (individual words) between output and reference text.
    \item ROUGE-2 measures the overlap of bigrams (two-word sequences) between output and reference text.
    \item ROUGE-L measures the longest common subsequence between output and reference text.
\end{itemize}

While small changes in temperature or weather conditions can lead to different predictions, making ROUGE not always the best metric for evaluating our weather forecast model, it has proven effective in predicting similarities between output and reference predictions. The presented scores indicate that the generated outputs have a good degree of similarity to the reference weather forecasts. We obtained F-scores of around 0.69 for ROUGE-1, 0.53 for ROUGE-2, and 0.68 for ROUGE-L. The F-score is the harmonic mean of the precision and recall values, where precision is the proportion of unigrams in the generated summary that appear in the reference summary, and recall is the proportion of unigrams in the reference summary that appear in the generated summary.

Another metric that we have deployed is BLEURT \cite{bleurt}, which evaluates the quality of Natural Language Generation. It is based on BERT, a transformer-based model that is pre-trained on a large corpus of text. BLEURT is a fine-tuned version of BERT that is trained to evaluate the quality of generated text, in terms of how well it captures language nuances and context. BLEURT analyses two sentences—a reference and a candidate—and scores how well the candidate sentence maintains the meaning of the reference and its fluency. We have employed BLEURT to assess the quality of our forecasts. Our evaluation dataset yielded a mean score of 0.6445. Each individual score takes values between 0 and 1, with values closer to 1 indicating a better similarity. Experimentally, values in the range of 0.7 indicate strong similarity, however, we have found that even minor numerical changes in the forecast can result in significantly different predictions, which BLEURT might not take into account in its evaluation.

To conclude, automated metrics such as BLEURT and ROUGE are fast and inexpensive to deploy, and are a good indicator on the LLMs ability to generate coherent text, thus ruling out bugs or other faulty aspects that may compromise the fine-tuning process. However, they cannot, by themselves, capture all aspects of the model's performance, particularly with respect to the weather forecast generation task.

\subsection{Human and GPT-4 Evaluation}
To get a good grasp on our model’s performance, we turned to human evaluation, to get a deep understanding of our model performance. We asked human evaluators to annotate our models' outputs in terms of how accurate the weather forecast is based on the input JSON. Each pair of input-output entries is assigned a score between 0 and 1, where 0 indicates that the output is not at all accurate, 0.5 indicates partial accuracy: the overall weather trend is well captured, however there are small differences between the JSON and the generated text, and 1 indicates full accuracy - the output is almost or completely correct and can be completely trusted. We average the human scores to get a final human evaluation score for our LLM.

We have obtained a mean score of 0.82 (Figure \ref{fig:fig10}), using human evaluation which illustrates good forecasting performance. This metric is consistent with BLEURT, but also much higher, since many forecasts may be phrased differently, e.g. enumerating weather traits such as wind, precipitation in different orders. Human evaluation was only possible for a representative sample of approx 500 dataset entries, due to it being time-consuming and expensive to conduct. However, we can approximate it on a larger scale, by employing a much larger language model (LLM), such as GPT-4 \cite{gpt4}, prompted to compare the output and reference text and provide a score of 0, 0.5 or 1, just as the human evaluators did. We use the following prompt for the GPT-4 model:

\begin{quote}
\texttt{For the following AI-generated weather bulletin, provide a ranking of how plausible it is given the weather data in the provided weather API JSON. A rank of 0 means completely not plausible, 1/2 means it is ok but has some mistakes, 1 means perfectly plausible. Your answer should be a single number.}
\texttt{<INPUT\_JSON>}
\texttt{<METEO\_BULLETIN\_TEXT>}
\end{quote}

The GPT-4 score of 0.81 is also high, indicating strong performance and a good alignment with human judgement, meaning this approach can be employed to approximate human evaluation in future evaluations for the LLMs involved in the MS platform.

\subsection{Results and Discussion}
\begin{figure}[htbp]
    \centering
    \includegraphics[trim={1cm 1cm 1cm 1cm}, clip, width=0.5\textwidth]{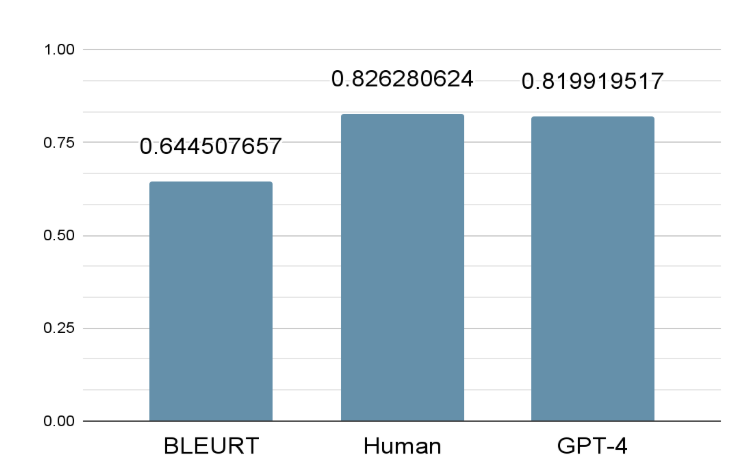}
    \caption{Model evaluation scores.}
    \label{fig:fig10}
\end{figure}

To conclude, we have deployed four different strategies to assess the performance of our LLM: ROUGE, BLEURT, coupled with human evaluation, as well as GPT-driven, prompt-based evaluation. The distilled results are illustrated in Figure \ref{fig:fig10}. All scores indicate a strong performance of our LLM. They also validate the fact that our AI-based approach is feasible as an architecture design and can be extended for the complete implementation of the MS platform.

\section{User Experience, Privacy, and Security}

\subsection{User Interface and Experience}
During the design phase of the MS platform, we focused on three key characteristics:
\begin{itemize}
    \item Users with no IT experience should be able to efficiently use the platform.
    \item All information should be accessible with a minimal number of clicks.
    \item Information should remain readable regardless of how the data is formatted, stored, or provided by Service Providers.
\end{itemize}

We quickly realised that Large Language Models (LLMs) could effectively meet these criteria. Consequently, we designed this MVP as a conversational interface. Incorporating natural language interaction significantly enhances the user experience by making the MS platform more intuitive and accessible for non-technical users. This approach allows users to interact with the system using everyday language, eliminating the need for specialised technical knowledge. As a result, users can easily find the information they need and perform tasks more efficiently, leading to a smoother and more enjoyable experience on the platform.

We incorporated several small details to enhance user experience, including reactive buttons that change colour when hovered over and show a pushing animation when clicked, a chat input text box that scrolls down as the user types to always display the most recent three lines of the query, a character limit of 1500 for input with a real-time character count to prevent overly long inputs, restrictions on sending empty messages with a brief red border signal, local caching of chat history until the page is closed or cleared, scrollable chat history, and a loading animation while waiting for responses with the “Send” button disabled during processing.

\subsection{Privacy and Security Measures}
Our MVP implementation safeguards against cross site scripting (XSS), SQL injection, and other common security vulnerabilities to ensure that the interface is secure and that user data is protected. We also follow best practices for data encryption to safeguard user information, aiming to use secure connections (HTTPS) and encrypt sensitive data at rest and in transit. Furthermore, we offer complete transparency regarding data collection: we do not store any user data beyond what is necessary to provide the search service, namely the user's input and the model's response.

\section{Future Work}

\subsection{Extending to Multiple Smart Services}
Currently, the MVP for the MS platform retrieves and presents data from a single service (the weather forecast API). We plan to expand this by integrating data from multiple Smart Services and adding capabilities to identify and selectively query such relevant services. We outline three different strategies for data integration that can be supported by the MS platform:

\begin{itemize}
    \item \textbf{Model fine-tuning} involves creating tailored adapters for each Smart Service Provider in the FUTURAL project. This approach builds on the core principle behind the MVP implementation. Based on data characteristics from Smart Services, we will build a dataset that can be used to modularly create LLMs for each SS, as illustrated in Figure \ref{fig:fig2}. Each fine-tuned adapter will be combined with the base LLM per need, and interrogated per need depending on the user query. It is also possible to build a single FUTURAL LLM, with the advantage that MS platform implementation would be much simpler, however scaling to arbitrary datasets from different providers is more challenging and may pose text generation performance challenges. Model finetuning, either via adapters per Smart Service or as a single iteration, has the downside that it requires human intervention, including data curation, in order to integrate data from a new service;
    \item \textbf{RAG (Retrieval-Augmented Generation)} \cite{rag} is an alternative technique that can be used to more seamlessly integrate new Smart Service data. RAG is a highly effective technique for providing users with specific information without requiring complex processing. It combines conventional databases or APIs with the generative capabilities of large language models (LLMs). In this approach, a user query is first used to generate specific database queries. The most relevant results from these queries are then integrated into the text generation process. For example, consider a user query about suitable crop hybrids available in a certain area, known for their productivity. If a Smart Service has a database of such hybrid species, it will query the database and combine the top five results with the user query as input for the LLM. The LLM will then generate a text response listing the suitable candidates. In this case, the LLM does not require special interpretation capabilities for the queried data, unlike our weather MVP implementation. We estimate that many Smart Services will produce data that can be directly accessed using RAG, without needing specific model fine-tuning techniques.
    \item \textbf{API-call integration} \cite{toolformer} allows creating models that can invoke third-party APIs during the text generation process, without the need to deterministically process and classify the input query. This technique stems from the observation that LLMs are unable to process location-based queries (to achieve this, a user location should be somehow integrated manually in the query), time-based queries, or simple tasks such as arithmetic computations. For instance, in user queries such as: “search for farm properties that support space at least 70 tomato specimens and 50 cucumber specimens, with 1 square metre space for the former and 0.7 square metres for the latter”, most LLMs will not correctly compute the minimal required surface of the land. API-call integration allows models to learn from examples and invoke external API-calls during text generation, in order to supply accurate information that is not part of the model itself. In our example, a properly-trained LLM will generate the expression 70*1 + 50*0.7, but instead of using word-probabilities to generate the response, will invoke a calculator API in order to obtain the response. API-call integration can be used in support of the MS platform to increase the query accuracy for users and at the same time, it can be used to train models how, and when to call or query external Smart Services, as well as what Services should be called. Thus, API-call integration can be leveraged as an alternative to our deterministic Input Processing Module, in order to automatically make decisions regarding which Smart Service should be invoked.
\end{itemize}

\subsection{Supporting Different Data Formats}
It is expected that Smart Services from the project will produce data of different types, not necessarily in textual form. The most likely scenario is that data can take the form of graphs, plots, maps and other specific visuals. We will accommodate such data in the search process, by leveraging existing LLM capabilities to read and process images, together with metadata which will be available from Service Providers, in order to capture the characteristic of each image. We plan to extend search support for any kind of visual data artefact, as part of the MS implementation process.

\subsection{Scalability and Third-Party Data}
In order to give users a seamless experience, the interface should be able to handle a large number of users simultaneously, this way we can ensure that users can interact with the model without any delays or interruptions. For that, a robust server is required that can handle multiple requests concurrently. We aim to use a cloud-based server (such as Openstack) to host the interface, which will allow us to scale up the resources as needed. This aspect will be crucial when the interface is deployed and used by a large number of users. but the exact details of the scalable deployment will be determined in the future stages of the project, after the initial prototype is developed and the number of smart services and users can be estimated.

We aim to replicate our LLM server on multiple instances, and use a load balancer to distribute the incoming requests among the instances. This way, we can ensure that the server can handle a large number of requests without any delays. We will also use a caching mechanism to store the responses to frequently asked questions, which will reduce the response time and improve the overall performance of the system.

Furthermore, we wish to replicate both backend and frontend on multiple instances in different regions to ensure that the interface is available to users worldwide, with minimal latency. This will also help in reducing the load on a single server and improve the overall performance of the system. By replication of our servers, we can ensure that the interface is highly available and can handle a large number of users simultaneously, at any time of the day. We will also monitor the performance of the servers and scale up the resources as needed, to ensure that the interface is responsive and efficient.

The MS platform is designed to accommodate new data sources that may appear during or after the project duration. It is possible that certain data sources cannot, either for technical reasons or other reasons, be processed via the LLM approach described previously. Also, it may be possible that certain datasets will be useful in their raw shape, for domain experts or application developers, as a resource for research or for application development. For this reason, we will also maintain a metadata repository that will index all available data and data sources that are available in the FUTURAL project, via the MS platform, and will allow direct access to such data. This repository will be subject to future MVP developments.

\section{Conclusion}
Our MVP implementation for the MS platform is a standalone, fully functional Smart Service designed to interpret weather data and provide natural language responses to weather forecast queries. This MVP serves as a single-service search platform with the primary goal of pilot-testing the concept of generative AI for user searches over heterogeneous data within the FUTURAL project. Additionally, it aims to experiment with various Large Language Models and adaptation techniques to enhance their generative performance. 

The implementation and evaluation results demonstrate the feasibility of our approach. Moving forward, the next phase involves expanding the platform to include additional services and specific data sets from the FUTURAL project, thereby increasing the diversity of available data. This expansion will enable more comprehensive testing and refinement of the system, ultimately enhancing its robustness and versatility in handling diverse queries and datasets.

\bibliographystyle{plain}
\bibliography{paper}

@misc{mistral,
  author = {{Mistral AI}},
  title = {Mistral: Open-Source Large Language Models},
  year = {2023},
  howpublished = {\url{https://www.mistral.ai/}}
}

@misc{mistral_function_calling,
  author = {{Mistral AI}},
  title = {Function calling capabilities},
  year = {2023},
  howpublished = {\url{https://docs.mistral.ai/capabilities/function_calling/}}
}

@misc{llama2,
  author = {Touvron, H. and others},
  title = {Llama 2: Open Foundation and Fine-Tuned Chat Models},
  year = {2023},
  howpublished = {\url{https://research.facebook.com/publications/llama2/}}
}

@inproceedings{attention,
  author = {Vaswani, A. and others},
  title = {Attention is all you need},
  booktitle = {Advances in Neural Information Processing Systems},
  volume = {30},
  year = {2017}
}

@inproceedings{transfer_learning,
  author = {Ruder, S. and others},
  title = {Transfer learning in natural language processing},
  booktitle = {Proceedings of the 2019 Conference of the North American Chapter of the Association for Computational Linguistics: Tutorials},
  pages = {15--18},
  year = {2019}
}

@inproceedings{transformers,
  author = {Wolf, T. and others},
  title = {Transformers: State-of-the-art natural language processing},
  booktitle = {Proceedings of the 2020 Conference on Empirical Methods in Natural Language Processing: System Demonstrations},
  pages = {38--45},
  year = {2020}
}

@inproceedings{peft,
  author = {Houlsby, N. and others},
  title = {Parameter-efficient transfer learning for NLP},
  booktitle = {Proceedings of the 36th International Conference on Machine Learning},
  pages = {2790--2799},
  year = {2019}
}

@article{lora,
  author = {Hu, E. J. and others},
  title = {LoRA: Low-Rank Adaptation of Large Language Models},
  journal = {arXiv preprint arXiv:2106.09685},
  year = {2022}
}

@article{qlora,
  author = {Dettmers, T. and others},
  title = {QLoRA: Efficient Finetuning of Quantized LLMs},
  journal = {arXiv preprint arXiv:2305.14314},
  year = {2022}
}

@inproceedings{rouge,
  author = {Lin, C. Y.},
  title = {ROUGE: A Package for Automatic Evaluation of Summaries},
  booktitle = {Text Summarization Branches Out: Proceedings of the ACL-04 Workshop},
  pages = {74--81},
  year = {2004}
}

@inproceedings{bleurt,
  author = {Sellam, T. and others},
  title = {BLEURT: Learning Robust Metrics for Text Generation},
  booktitle = {Proceedings of the 58th Annual Meeting of the Association for Computational Linguistics},
  pages = {7881--7892},
  year = {2020}
}

@article{gpt4,
  author = {{OpenAI}},
  title = {GPT-4 Technical Report},
  journal = {arXiv preprint arXiv:2303.08774},
  year = {2023}
}

@inproceedings{rag,
  author = {Lewis, P. and others},
  title = {Retrieval-augmented generation for knowledge-intensive nlp tasks},
  booktitle = {Advances in Neural Information Processing Systems},
  volume = {33},
  pages = {9459--9474},
  year = {2020}
}

@article{toolformer,
  author = {Schick, T. and others},
  title = {Toolformer: Language models can teach themselves to use tools},
  journal = {arXiv preprint arXiv:2302.04761},
  year = {2023}
}

@article{vldb2023,
  author = {Arsan, Z. M. and others},
  title = {Large Language Models for Data Integration},
  journal = {Proceedings of the VLDB Endowment},
  volume = {16},
  number = {12},
  pages = {3799--3802},
  year = {2023}
}

@article{arxiv2023,
  author = {Sun, Z. and others},
  title = {A Survey on Large Language Models for Data Integration},
  journal = {arXiv preprint arXiv:2311.08583},
  year = {2023}
}

\end{document}